\documentclass[12pt,preprint]{aastex}

\usepackage{natbib}
\bibpunct[,]{(}{)}{;}{a}{,}{,}   

\newcommand{\HI}{\mbox{\sc      H{i}}}

\newcommand{\kms}{\mbox{km    s$^{-1}$}}

\slugcomment{The Evolving ISM in the Milky Way \& Nearby Galaxies}

\shorttitle{Spitzer Survey of Virgo}
\shortauthors{Kenney et al.}

\begin{document}


\title{Spitzer Observations of Environmental Effects on Virgo Cluster Galaxies}


\author{
Jeffrey D. P. Kenney\altaffilmark{1},
O.\ Ivy Wong\altaffilmark{1},
Anne Abramson\altaffilmark{1},
Justin H. Howell\altaffilmark{2},
Eric J. Murphy\altaffilmark{2} and
George X. Helou\altaffilmark{2}}

\affil{$^{1}$Astronomy Department, Yale University, P.O. Box 208101
New Haven, CT 06520-8101}
\affil{$^{2}$Spitzer Science Center, California Institute of Technology,
Pasadena, CA 91125}

\email{jeff.kenney@yale.edu}


\begin{abstract}

We present initial results from SPITSOV, the Spitzer Survey of Virgo,
which includes MIPS and IRAC observations for a carefully selected
sample of 44 Virgo cluster spiral and peculiar galaxies.
SPITSOV is part of a multiwavelength campaign to  understand the effects
of the cluster environment on galaxy evolution.
These Virgo galaxies are different from galaxies outside of clusters,
since most of them have been significantly modified by their environment.
The SPITSOV data
can serve the community as the Spitzer sample of nearby cluster spiral galaxies,
a complement to the SINGS data for nearby non-cluster galaxies.

In this paper we describe the sample, the goals of the study, and present
preliminary results
in 3 areas:
1. Evidence for ram pressure-induced disturbances in radio morphologies
based on changes in the FIR-radio (70$\mu$m-20cm) correlation.
2. Evidence for ram-pressure stripped extraplanar gas tails from comparisons
of dust/PAH (8$\mu$m) emission and optical dust extinction.
3. Evidence for unobscured star-forming regions with large ratios of H$\alpha$
to 24$\mu$m emission in some galaxies due to ram pressure stripping of dust.

\end{abstract}



\section{Introduction}

Studies of galaxies near and far show broad evidence for the basic picture of
hierarchical galaxy formation, in which large galaxies assemble from smaller
pieces, and interactions play a large role in evolution.
Yet the dominant interaction processes, and what actually happens
in them, are still not sufficiently well-known.
Many of the processes driving galaxy
evolution can be clearly observed in and around clusters.
Since hierarchical galaxy formation is scale-free, cluster formation is much like
galaxy formation
and the processes of gravitational interactions, gas accretion, and gas
stripping
occur both as galaxies fall into clusters and as small galaxies fall into larger
galaxies.
Thus by studying cluster galaxies we learn not only about their own evolution
but also about the processes which occur during galaxy formation.

The morphology-density relation \citep{dressler80} and the observed differences
between
clusters at different redshifts \citep{dressler97} show that the environment
significantly drives the evolution of galaxies. 
It is still not clear which are the dominant environmental processes in
clusters,
although both gravitational and gas dynamical processes seem to be important.
Gravitational effects include slow galaxy-galaxy interactions and
mergers \citep{mihos04}, harassment \citep{moore96,moore98},
the  effects of the global tidal field \citep{byrd90,henriksen96}
and the effects of sub-cluster merging
\citep{bekki99,stevens99}. Gas dynamical effects include ram pressure
and turbulent viscous stripping of the cold ISM
\citep{gunn72,abadi99,schulz01,vollmer01} and starvation \citep{larson80}.
While there is much new evidence for both gravitational interactions
and ram pressure gas stripping in cluster galaxies
\citep{mihos05,moran07,chung07,tonnenson07},
it remains a challenge to clearly identify which process is occuring
and to quantify the impacts for each type of interaction.

As the nearest rich cluster, Virgo is one of the best places to study
evolutionary processes in action.
It is close enough to show observational details that, when combined with
simulations,
can constrain key interaction parameters.
The majority of its galaxies have clearly been modified by the environment.
Virgo has a dense ICM, and has already provided some of
the best evidence for ongoing ram pressure stripping. The cluster
also exhibits a lot of substructure, with infalling groups containing
tidally interacting galaxies and mergers.
There is evidence for ongoing sub-cluster cluster mergers
\citep{schindler99}, allowing
us to study the possible impacts of a dynamic ICM and a varying global
potential. There are also $\HI$-deficient galaxies far away from the dense
core. These are ideal candidates to study the mysterious mechanisms
that seem to affect galaxies far beyond the virial radius
\citep{solanes01,balogh98}.

Here we present the Spitzer Survey of Virgo (SPITSOV), an infrared
imaging study of Virgo Cluster galaxies using the IRAC and MIPS
instruments aboard the Spitzer Space Telescope. SPITSOV forms part of our large
multi-wavelength study of Virgo cluster
galaxies, which includes optical BVRH$\alpha$, HI, 20cm radio continuum, GALEX
UV, and optical spectroscopy.
A major goal of our studies is to identify clear diagnostics for each process
from detailed studies of individual galaxies,
and to quantify the impacts for each type of interaction.
Our goal for the Spitzer study is to compare IR distributions with other
wavelengths
and with
numerical models, to improve our understanding of the ISM, star formation, and
stellar populations
in environmentally-modifed cluster galaxies.

In this paper we briefly describe the galaxy sample,
the data processing, and three examples of how the SPITSOV data
is used to learn about ram pressure stripping in clusters.
Many basic things are still not known about stripping, such as:
How does the real, complex, multi-phase ISM react to ram pressure?
How do distinct parts of the ISM react?
Are shocks driven into the unstripped ISM by the strong external pressure?
Under what circumstances is star formation enhanced or disrupted?
What is the fate of star-forming molecular clouds?
What are the rates of triggered star formation vs. gas removal?
Does stripping happen only in the core, or does it also occur in the outskirts,
perhaps due a dynamic rather than a static ICM?
Spitzer data allows us to address many of these questions.

\section{The sample}

\begin{figure*}
\begin{center}
\includegraphics[scale=0.63]{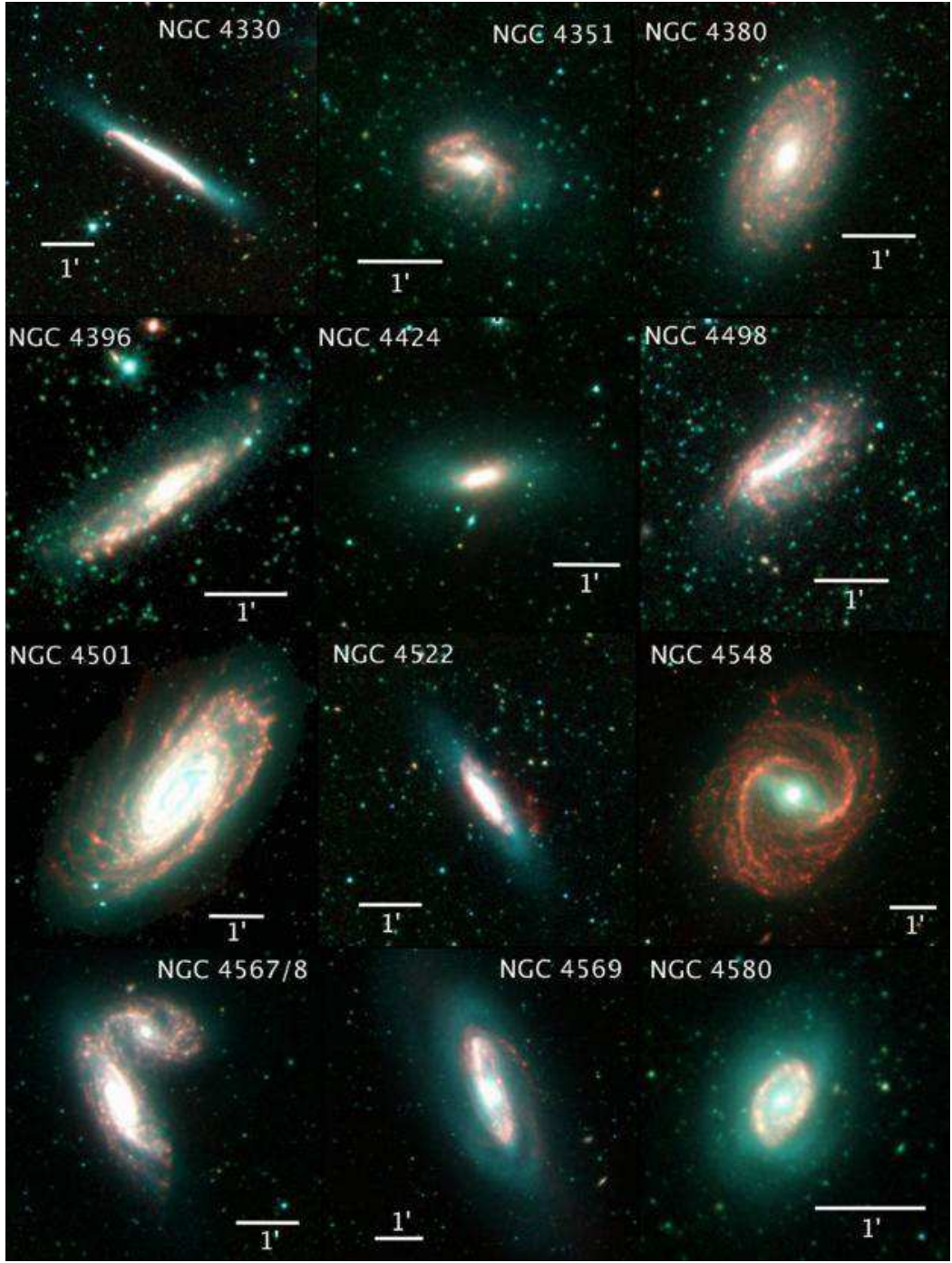}
\end{center}
\caption{\footnotesize{Three-color IRAC composite images of twelve Virgo cluster
galaxies from the SPITSOV sample.  Blue and green represent the
3.6 $\mu$m and 4.5 $\mu$m emission, respectively, and the 5.8 $\mu$m
and 8.0 $\mu$m emission are shown in red.
Many Virgo spirals lack gas and PAH emission in their outer disks
due to ram pressure stripping. Some of these are in an active stripping
phase and have extraplanar gas, PAH emission and star formation
(NGC 4522, NGC 4569, and NGC 4330).
}}
\label{compendium}
\end{figure*}

We have chosen a representative sample of 44 S0/a-Sm Virgo spirals,
with a range of 20 in mass, a range of H$\alpha$ and HI properties,
and spread throughout the cluster.
With this sample we can explore the range of galaxy interactions
and evolutionary stages which exist in Virgo spirals,
and conduct statistical analyses.
The SPITSOV sample is the MIR-FIR counterpart to the H$\alpha$
\citep{koopmann04a} and HI VIVA
(VLA Imaging of Virgo in Atomic gas) Virgo cluster samples
\citep{chung07,chung08}.

The spatial distribution of the survey galaxies
has good sampling from the core to the cluster outskirts, including
galaxies both within and beyond the X-Ray emitting region.
Since there is much recent evidence showing significant evolution
in the outskirts of clusters \citep{solanes01,balogh98},
we have included galaxies located out as far as 10$^{\circ}$ (=2.5 Mpc=3R$_{\rm
vir}$).
The galaxies have a wide range of Hubble types and H$\alpha$ distributions
including Normal, Anemic, Starburst, and Truncated \citep{koopmann04b}.
We believe that these categories represent different types
of interactions, different evolutionary phases, and the effects of
different parameters for a given type of interaction. Figure~\ref{compendium}
shows
three-color composite IRAC images of twelve
galaxies from the sample.  These include galaxies with normal stellar disks and
evidence for ongoing or past ICM-ISM stripping,
some with disturbed stellar disks indicating gravitational encounters,
and some with peculiarities that are not yet understood.


\section{Observations}
Each of our Virgo Cluster galaxies are fully-mapped using the IRAC and MIPS
imagers.  The IRAC  3.6 $\mu$m and 4.5 $\mu$m images trace the stellar populations,
while the IRAC 8 $\mu$m images map the PAH emission.  The MIPS
24 $\mu$m is a good indicator of unobscured star formation and the MIPS
far-infrared bands (70 $\mu$m \& 160 $\mu$m) probe the cooler dust population
which accounts for the majority of the total dust mass.

Our observing strategy is guided by that of the SINGS Legacy team
\citep{kennicutt03}.  However, our observing integration times are twice that
of SINGS because  we are interested in detecting extraplanar and outer
galaxy infrared emission.  Our integration times correspond to 3-$\sigma$
sensitvity levels of 0.014, 0.020, 0.077, 0.089, 0.14, 0.55 and 2.39 MJy/sr at
3.6, 4.5, 5.8, 8.0, 24, 70 and 160 $\mu$m, respectively.

The basic calibrated datasets (BCDs) delivered by the Spitzer Science Center
(SSC)
are further processed using the standard Post-BCD tools developed by the SSC.
The SSC data reduction pipeline is also tuned to achieve further improvement
in the data (especially in the 70 \& 160 $\mu$m images).

\section{Results}
\subsection{Evidence for Ram Pressure from Changes in the FIR-Radio Correlation}

Although $\HI$ observations can show whether a galaxy was stripped, radio
continuum emission can be used to differentiate between active and past
stripping.
A clear tracer of ram pressure
can help show whether the galaxies are stripped in the cluster cores or in the
outskirts
(which seems to occur for some galaxies).
Is ram pressure sometimes stronger
than that expected for a smooth static ICM? If the ICM is dynamic and lumpy,
as might be expected during subcluster mergers, this can make it easier
to strip galaxies.  Thus clear tracers of ram pressure are relevant for learning
the relative importance of stripping to other processes which drive galaxy
evolution.

\begin{figure*}[h]
\begin{center}
\includegraphics[scale=0.5]{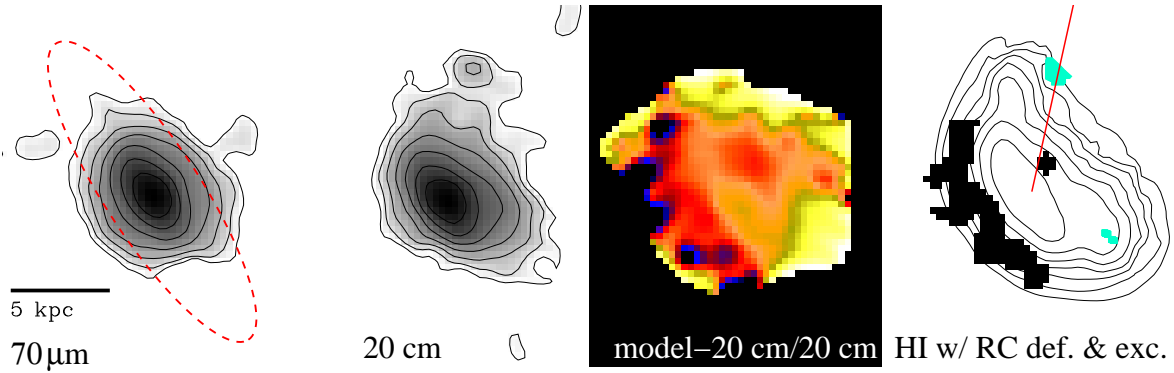}
\end{center}
\caption{\footnotesize{
Radio and far-infrared maps of Virgo spiral NGC 4522, known to be
experiencing ram pressure stripping.
a.) MIPS 70$\mu$m map.
b.) VLA 20cm radio map.
c.) Ratio of smoothed 70$\mu$m (=``expected radio'') to observed 20cm radio
emission.
d.) HI contours with radio deficit regions marked in black and radio excess
regions
marked in cyan.
The radio deficit region is where the observed radio brightness is less
than 50\% of the expected value, based on the 70$\mu$m map \citep[from ][]{ 
murphy08b}.
}}
\label{rc+fir}
\end{figure*}

We find clear indications of ram pressure from a comparison of radio and
far-infrared distributions, in a preliminary sample of 10 Virgo galaxies \citep{murphy08b}.  Figure~\ref{rc+fir} shows maps for the Virgo spiral NGC
4522.
Both the radio and FIR distributions (along with all ISM tracers) are truncated
at 0.4R$_{25}$,
as the outer ISM has been stripped from the galaxy, leaving an ISM-free outer
stellar disk \citep{kenney04}.
However the radio and FIR have different distributions. Whereas the FIR map is
relatively symmetric, the radio map shows clear asymmetries, with compressed
contours to
the southeast, and an extended tail to the northwest. The radio-emitting cosmic rays
and
magnetic fields are more strongly affected by ram pressure than the denser
components
of the ISM, as traced by the  70$\mu$m emission.

Spitzer 70$\mu$m maps help show evidence for ram pressure by providing a way of
calibrating ``normalcy" for the easily disturbed radio continuum distribution.
We use the Spitzer 70$\mu$m maps and the FIR-radio correlation to predict
what the radio continuum distribution would be in the absence of ICM pressure.
Differences between the observed and predicted radio maps show ``radio deficit"
regions near the leading edges of several galaxies, and ``radio excesses"
on the trailing side, clearly indicating ongoing ICM-ISM interactions.
Galaxies with local radio deficits seem to have global radio enhancements
(relative to the FIR emission), suggesting that ICM-ISM interactions
accelerate cosmic ray electrons \citep[see ][]{murphy08a,murphy08b}.

\begin{figure*}
\begin{center}
\includegraphics[scale=0.52]{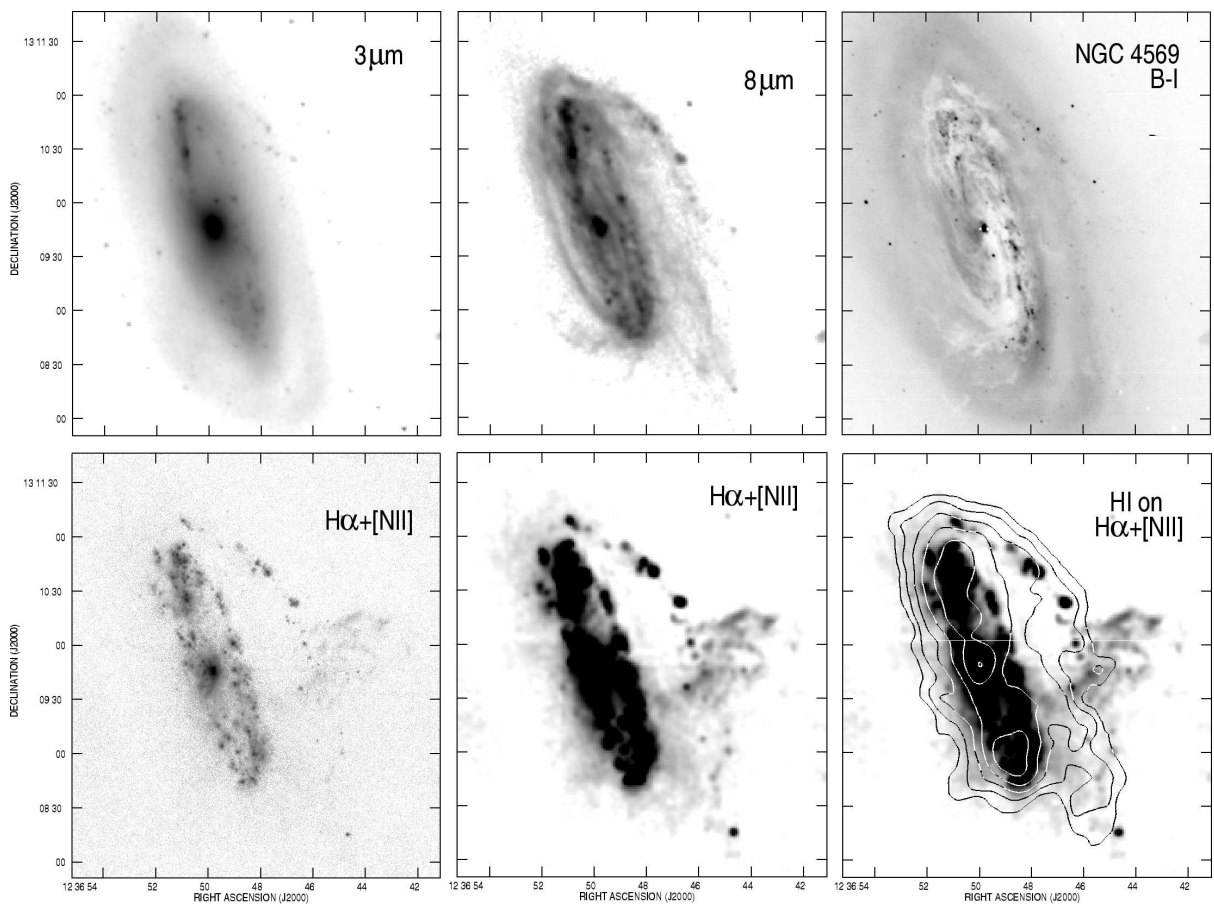}
\end{center}
\caption{\footnotesize{
Six-panel montage of central 4$'$=16 kpc in NGC 4569, a large
Virgo spiral experiencing ram pressure stripping.
a.) Spitzer IRAC 3.6$\mu$m showing stellar distribution.
b.) Spitzer IRAC 8$\mu$m showing PAH/dust emission.
c.) WIYN B-I image showing dust extinction as white (``red'')
young star clusters as black (``blue'').
d.) H$\alpha$+[NII] image, showing star formation truncated in disk at
1.4$'$=6.5 kpc, and arm of extraplanar HII regions in West.
e.) Deeper, smoothed H$\alpha$+[NII] image, showing diffuse extraplanar
nebulosity from nuclear outflow, as well as arm of extraplanar HII regions.
f.) HI contours on H$\alpha$+[NII] greyscale image, showing extraplanar
$\HI$ arms extending to the West.
Note that the extraplanar emission is detected with very good sensitivity
and resolution at 8$\mu$m.
}}
\label{n4569-6panel}
\end{figure*}

\begin{figure*}
\begin{center}
\includegraphics[scale=0.8]{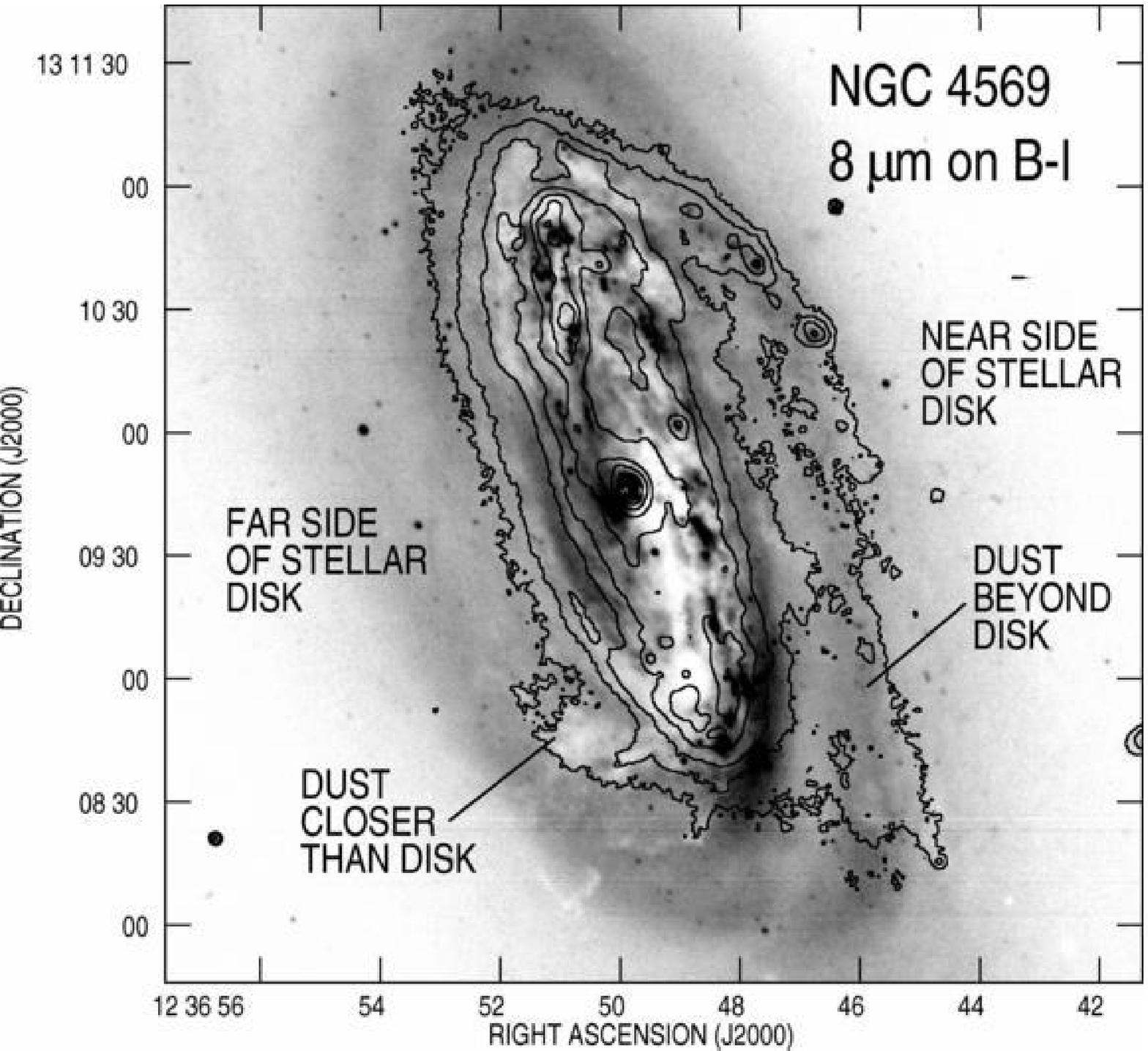}
\end{center}
\caption{\footnotesize{
IRAC 8$\mu$m contours on WIYN B-I greyscale image
in the Virgo spiral NGC 4569.
Dust within the disk plane produces strong
optical extinction on the near side of the stellar disk,
which is to the west.
The anomalous western  complex show strong
8$\mu$m/PAH/dust emission but weak optical dust
extinction (``red'' B-I colors, shown as white),
indicating that this gas and dust lies behind the stellar disk
and is therefore extraplanar.
The southeastern extraplanar feature shows weak
8$\mu$m/PAH/dust emission but strong
extinction, indicating that this
gas and dust lies in front of the stellar disk.
}}
\label{n4569-8um+b-i}
\end{figure*}

\subsection{Evidence for Extraplanar Gas Tails from Comparisons of 8$\mu$m and
B-I
Images}

One-sided extraplanar gas distributions in galaxies with undisturbed
stellar disks are clear signatures of ram pressure stripping.
They can be easily identified in nearly edge-on galaxies, but
are harder to identify in less inclined galaxies like NGC 4569.
In such galaxies, comparisons of 8$\mu$m PAH images with B-I
``dust extinction'' maps can demonstrate that gas is extraplanar, and thereby
constrain interaction models by clarifying the ISM geometry.

One of the largest Virgo spirals with evidence for ongoing
ICM-ISM stripping is NGC 4569, in which
all disk ISM tracers (HI, 8$\mu$m, H$\alpha$) are sharply truncated at
30\% of the optical radius R$_{25}$, as shown in Figure~\ref{n4569-6panel}.
Anomalous arm-like features of 8$\mu$m, HI and H$\alpha$ to the west
of the truncated disk may be gas stripped from the disk by ram pressure.
This has been suggested for the HI by \citep{vollmer04},
however the gas morphology and
kinematics alone do not demonstrate that the gas is extraplanar.

A comparison of the relatively strong 8$\mu$m dust/PAH emission with
the relatively weak associated dust extinction in the B-I image,
shown in  Figure~\ref{n4569-8um+b-i}, indicates that
the anomalous western dust and gas must lie $\it{behind}$ the stellar disk.
The extinction associated with this much dust would be much larger if this
dust were within or in front of the disk, as seen by a comparison with the
extinction
in the inner galaxy.
This clearly shows that the anomalous western gas and dust are extraplanar
and therefore have been stripped from the disk.
Moreover, it shows that this extraplanar gas is on the far side of the disk,
as expected for stripping from a galaxy which is moving through the
intracluster medium towards us. NGC 4569 is one of the few blueshifted
galaxies, with a line-of-sight velocity of -235 $\kms$, and its orbit through
the
cluster must have a large component toward us.

While the western extraplanar gas feature confirms the
basic stripping scenario, there is another ISM feature which doesn't fit in
with the simplest picture. A small feature in the
southeast beyond the gas disk truncation radius is clearly observed in the
8$\mu$m, \HI,
and H$\alpha$ emission even though it is weaker and has less mass than the
western
extraplanar feature (see Figure~\ref{n4569-6panel} and
Figure~\ref{n4569-8um+b-i}).
However the dust extinction traced by the B-I image from this feature
is strong, as seen by a comparison with the extinction in the
inner galaxy on the same side of the major axis.
This indicates that the southeast feature is also extraplanar,
but unlike the much larger western extraplanar gas feature,
it lies on the near side of the stellar disk.
How is this to be understood, in the context of a scenario
in which stripping acts to push gas to the far side of the galaxy?

The explanation may be gas fallback after peak pressure.
Comparisons between
simulations and observations \citep{vollmer04} as well as stellar population
studies \citep[showing that star formation has stopped ~300 Myr ago; ][]{crowl08} indicate
that
NGC 4569 is currently observed ~300 Myr after peak pressure.
Simulations show that after peak pressure, some gas which had been
pushed upwards by ram pressure falls back to the galaxy.
Perhaps the SE extraplanar feature is such a fallback feature.
Regardless of the explanation, it is clear that
comparisons of 8$\mu$m PAH emission with B-I ``dust extinction'' images
provide valuable constraints on interaction models by clarifying the ISM
geometry \citep[more details can be found in][]{kenney08}.

\subsection{Star Formation: Triggered, Obscured, or Revealed}

Among the biggest impacts of interactions on galaxy evolution
are changes in the rates and locations of star formation.
In clusters, there are many things that can alter galaxies' star formation
histories
that we would like to understand.
Is star formation triggered by gravitational interactions or ram pressure?
For galaxies experiencing ram pressure,
what are the relative rates of gas loss due to stripping and triggered star
formation?
How much star formation occurs in galaxy disks versus galaxy halos
and intracluster space?
Studies to date have attempted to address these questions with  H$\alpha$
imaging \citep{koopmann04a,koopmann04b},
but the results are somewhat compromised by extinction effects.
We wish to explore these questions
by using the MIPS 24$\mu$m dust emission in combination with the H$\alpha$
emission \citep{calzetti07}, to derive more accurate
star formation rates and distributions in cluster galaxies.

\begin{figure*}[h]
\begin{center}
\includegraphics[scale=0.53]{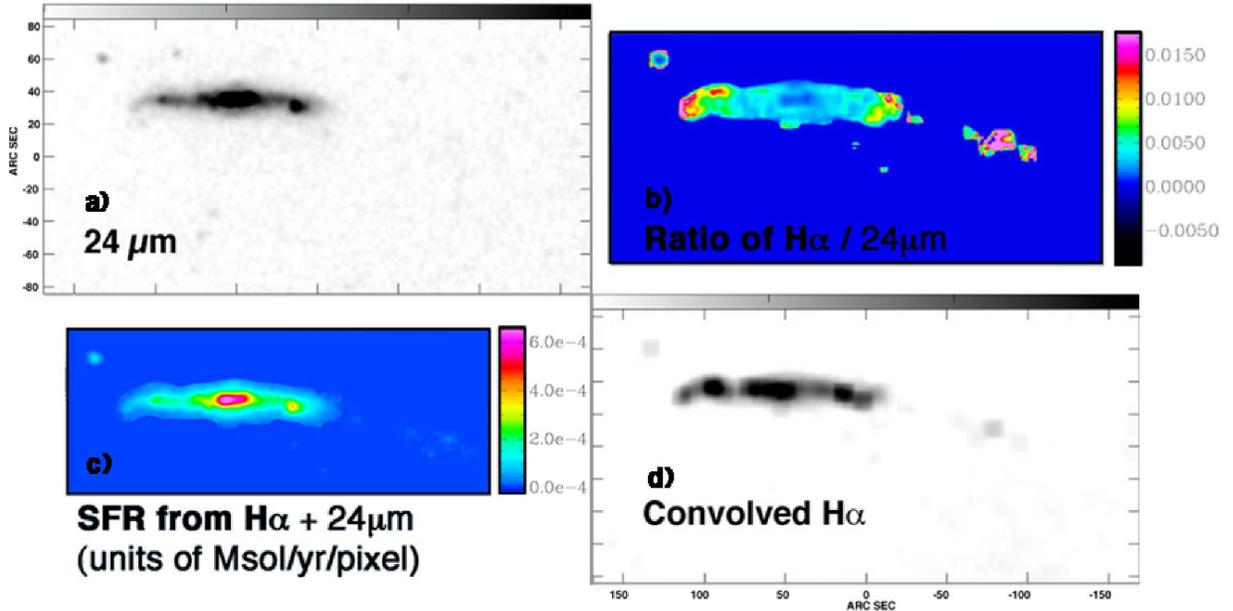}
\end{center}
\caption{\footnotesize{
Star formation images of NGC 4330, experiencing ram pressure.
a.) MIPS 24$\mu$m image.
b.)  H$\alpha$/24$\mu$m ratio map, showing the star forming regions
which are most and least obscured by dust. Regions near the
leading edge and the extraplanar tail are least obscured.
c.) Combined star formation map, a linear combination of
24$\mu$m and H$\alpha$ images.
d.) WIYN H$\alpha$ image, convolved to same 6$''$ resolution as 24$\mu$m map.
}}
\label{n4330-4panel}
\end{figure*}

There is suggestive evidence based on H$\alpha$ imaging
for triggered star formation due to ram pressure
in some Virgo spirals \citep{koopmann04b},
but this was not conclusive due to dust extinction effects.
For example the highly-inclined Virgo spiral NGC 4402,
which is clearly experiencing ICM ram pressure,
has optically luminous star-forming regions at its ``leading edge''
\citep{crowl05}.
Yet it is unclear whether the ICM wind has triggered star formation,
or merely removed some of the dust thereby exposing ``normal'' star formation.
Preliminary results from combining the 24$\mu$m with H$\alpha$ emission
suggest only modest enhancements in star formation
triggered by ram-pressure \citep{wong08,abramson08},
although the full SPITSOV sample needs to be studied,
and star formation traced by UV emission may also need to be incorporated.

Do interactions cause star formation in cluster galaxies to be obscured
more or less than non-cluster galaxies?
Preliminary results in a few galaxies indicate that ram pressure stripping may
remove dust from some
star forming regions, allowing more of the optical and UV photons to escape.
Figure~\ref{n4330-4panel} show
star-forming regions with high  H$\alpha$/24$\mu$m ratios
at the leading edge and the extraplanar tail
of the galaxy NGC~4330.
This could prove to be an indicator of active pressure on the galaxy,
and may provide interesting targets for detailed optical/UV studies of
unobscured young star clusters.

\section{Summary}

The Spitzer Survey of Virgo (SPITSOV) is an imaging study of 44 Virgo
Cluster spiral galaxies using the Spitzer Space Telescope,
whose goal is to advance our understanding of the
environmental effects on cluster galaxy evolution.
We describe preliminary results in 3 areas.

Spitzer 70$\mu$m maps and the FIR-radio correlation are used to predict
what the radio continuum distribution would be in the absence of ICM ram
pressure.
Differences between the observed and predicted radio maps show ``radio deficit"
regions near the leading edges of several galaxies, and ``radio excesses"
on the trailing side, clearly indicating ongoing ICM-ISM interactions.
Radio deficit regions may be an excellent diagnostic of active ram pressure.
The galaxies with local radio deficits seem to have global radio enhancements,
perhaps due to the acceleration of cosmic rays by shocks driven by ram pressure.

We combine the H$\alpha$ and 24$\mu$m images to produce star formation maps
and explore the effects of interactions on star formation.
Preliminary results in a few galaxies
suggest that star formation is only moderately enhanced
by ram pressure. Star-forming regions with large ratios of H$\alpha$ to 24$\mu$m
emission
are observed at the edges of some galaxies, probably due to
ram pressure sweeping of the dust away from star forming regions.
Large H$\alpha$/24$\mu$m ratios at galaxy edges may be a good
diagnostic of ongoing ram pressure.

For galaxies experiencing stripping, comparisons of dust emission and extinction
can
constrain interaction models by clarifying the 3D ISM distribution.
IRAC 8$\mu$m and optical B-I (``dust extinction'') images are used to
show whether the dust is in front of or behind the stellar disk.
In the case of the Virgo spiral NGC 4569, this comparison shows that the
``anomalous" western dust and gas features lie behind the stellar disk and are
clearly
extraplanar, and therefore have been stripped from the disk.

\acknowledgments
This work is based in part on observations made with the Spitzer Space
Telescope, which
is
operated by the Jet Propulsion Laboratory, California Institute of Technology
under a contract with NASA. Support for this work was provided by NASA through
an award issued by JPL/Caltech.




\end{document}